# An improved bearing fault detection strategy based on artificial bee colony algorithm

Haiquan Wang [1], Wenxuan Yue [2], Shengjun Wen[1], Xiaobin Xu[3], Menghao Su[2], Shanshan Zhang[2]，Panpan Du[2]

[1] Zhongyuan Petersburg Aviation College, Zhongyuan University of Technology, 41 Zhongyuan Road, Zhengzhou 450007, China
[2] Faculty of Electrical and Engineering, Zhongyuan University of Technology, 41 Zhongyuan Road, Zhengzhou 450007, China
[3] School of Automation, Hangzhou Dianzi University, 115 Wenyi Road, Xihu District, Hangzhou 310018, China
* Correspondence: wanghq@zut.edu.cn; Tel: 13523432980

**Abstract:** The operating state of bearing directly affects the performance of rotating machinery and how to accurately and decisively extract features from the original vibration signal and recognize the faulty parts as early as possible is very critical. In this study, the one-dimensional ternary model which has been proved to be an effective statistical method in feature selection is introduced and shapelets transformation is proposed to calculate the parameter of it which is also the standard deviation of the transformed shaplets that is usually selected by trial and error. Moreover, XGBoost is used to recognize the faults from the obtained features, and an improved artificial bee colony algorithm(ABC) where the evolution is guided by the importance indices of different search space is proposed to optimize the parameters of XGBoost. Here the value of importance index is related to the probability of optimal solutions in certain space, thus the problem of easily falling into local optimality in traditional ABC could be avoided.The experimental results based on the failure vibration signal samples show that the average accuracy of fault signal recognition can reach 97% which is much higher than the ones corresponding to other extraction strategies, thus the ability of extraction could be improved. And with the improved artificial bee colony algorithm which is used to optimize the parameters of XGBoost, the classification accuracy could be improved from 97.02% to about 98.60% compared with the traditional classification strategy.

## 1. Introduction

Rolling element bearings, as high-precision components that allow the rotating machine to run at extremely high speed, have been widely used in various precision instruments [1]. But due to the influence of overheating or overloading as well as the friction between different parts of the mechanical bearing, different types of failures often occur which eventually cause motor failures, lead to high maintenance costs, serious economic losses and safety hidden dangers [2]. Therefore, the diagnosis of bearing fault types and the prediction of the effective life of the machine are of great significance [3-4], Since the accurate bearing fault diagnostics has been approached by developing a physical model of bearing faults, and understanding the relationship between bearing faults and measurable signals which can be captured by a variety of sensors and analyzed with signal processing techniques, and feature extraction should be executed firstly for identifying the most discriminating characteristics in signals. Actually, several traditional time-domain or frequency-domain or time-frequency domain analysis methods such as spectral analysis [5,6], cyclostationary approach [7], fast Fourier transform[8-9], wavelet transform[10] and so on are used to extract useful features from signals but various questions exist unavoidably as the complex working environment and signal formats. Melih Kuncan et al. proposed a new vibration signal feature extraction method named One-dimensional ternary pattern (1D-TP) [11] which could transform from the idea of image feature extraction. The method could fully describe the oscillation as well as the flat parts in all obtained signals and it can be extracted effectively in real time if the fault frequency of the signal changes. However, the establishment of ternary pattern requires selecting the parameters of center point $P$ and threshold $\beta$ which are often determined by trial and error due to the influence of the abnormal value of the signal itself. In this paper, with the help of shapelets transform [12-13] which is a method of extracting similar subsequences, $P$ and $\beta$ are calculated as the center of the shapelets and the standard deviation of shapelets repectively. As the shaplets transform has the strong ability of denoising, the accuracy of feature extraction could be also improved.

Based on the extracted features, an effective classification strategy needs to be introduced such as XGboost constructed with boosting trees, which has proven their superiority in many classification applications [14-15]. But the selection of parameters of the XGboost, such as the learning rate and the number of weak classifiers, always relied on experiences, and the performance of the classifier can't achieve the optimization. In order to improve its performance, more and more optimization algorithms are introduced. Chen et al. proposed a combined prediction method based on LSTM and XGBoost [16], Zhang et al. proposed the Genetic Algorithm_XGBoost model, which used the optimization ability of genetic algorithm to perform multiple searches on the parameter combination of XGBoost to obtain a near-optimal solution[17], the improved Particle Swarm Optimization algorithm is used to optimize the XGBoost parameters, and established a machine learning model to predict the tensile strength and plasticity of steel[18]. In this paper, we concentrate on artificial bee colony algorithm, developed by Karaboga [19] based on simulating the foraging behavior of honey bee swarm, where the numerical comparisons demonstrated that the performance of ABC algorithm is competitive to other population-based algorithms with an advantage of employing fewer control parameters [20–22]. But up to now, there isn't any reference about the application of ABC for optimizing XGBoost or other ensemble learning algorithm. What is more, ABC algorithm can easily get trapped in the local optima when solving complex multimodal problems [22]. Thus in this paper ABC algorithm is creatively introduced to optimize XGBoost classifier, and aiming to escape from local optimal, an improved bee colony algorithm where the search is guided with the importance indices of sub-regionof solution space is proposed.

The rest of this paper is organized as follows. In Section 2, a brief introduction about thebearing fault is provided . Then a new feature extraction method combined with 1D-TP feature extraction method and shaplets transform is proposed in Section 3. In Section 4, in order to improve the fault detection accuracy, an improved artificial bee colony algorithm is created and applied to the design of XGBoost classifier. Relative experiments are exectued in Section 5 and the performance of the proposed feature extraction method and the improved ABC-XGboost are verified. Finallythe content of the paper is summarized and analyzed, and the future research directions is briefly described.

## 2. Bearing fault description

A rolling-element bearing consists of the outer race typically mounted on the motor cap, the inner race to hold the motor shaft, the balls or the rolling elements, and the cage for restraining the relative distances between adjacent rolling elements. As the most vulnerable component in motor drive system, four types of misalignments which are misalignment, shaft deflection, tilted outer race and tilted inner race are likely to cause the bearing failures.

As shown in Table 1, the data samples constructed for different defect sizes under different fault types (normal signal, inner ring fault, ball fault, outer ring fault) are listed. Single point faults are introduced to the bearings under test using electro-discharge machining with fault diameters of 7 mils, 14 mils, 21 mils, 28 mils and 40 mils, at the inner raceway, the rolling element and the outer raceway. Vibration data are collected for motor loads from 0 to 3 hp and motor speeds from 1,720 to 1,797 rpm using two accelerometers installed at both the drive end and fan end of the motor housing, and sampling frequency of 48 kHz were used.

**Table 1.** Establishment of sample data

| Bearing condition | Fault size | label | Training set | Test set |
|---|---|---|---|---|
| NS | 0 | 0 | 450 | 150 |
| IRF | 0.007 inch | 1 | 450 | 150 |
| | 0.014 inch | 2 | 450 | 150 |
| | 0.021 inch | 3 | 450 | 150 |
| BF | 0.007 inch | 4 | 450 | 150 |
| | 0.014 inch | 5 | 450 | 150 |
| | 0.021 inch | 6 | 450 | 150 |
| ORF | 0.007 inch | 7 | 450 | 150 |
| | 0.014 inch | 8 | 450 | 150 |
| | **0.021 inch** | **9** | **450** | **150** |

## 3. Feature Extraction

### 3.1. One-dimensional ternary patterns

1D-TP is develped based on the local ternary pattern [23-24], which is generally used in image processing for texture analysis. Different from directly processing multi-dimensional matrices in local ternary mode, 1D-TP uses patterns obtained from the comparisons between two neighbors of the vibration signals. The procedure of algorithm could be described as follow:

After defining relative parameters, such as the center point in the transformed vector $P_c$, the point $P_i$ within a specific range around $P_c$, as well as the range threshold $\beta$ which affects the encoding result, the value of the member within the threshold range $\beta$ is compared to neighbors in 1D-TP with Eq.1, and then ternary patterns code could be generated.

$$binarycode = \begin{cases} 1 & P_c > P_i + \beta \\ 0 & P_c \in (P_i + \beta, P_i - \beta) \\ -1 & P_c < P_i - \beta \end{cases} \qquad (1)$$

Based on the encoding results, the positive and negative values are separated and form two columns of binary codes, and converted to a decimal number who will replace the center value with the generated one. Thus two different sets of feature vectors could be obtained.

The 1D-TP method can fully describe the peak signal as well as the flat part in the vibration signal which considers the integrity of the signal. Although it has been proved to be effective in extracting features from raw data, there are still two problems: the first one is how to choose the value of parameter β which is often obtained by trial and error to guarantee the best extraction effect. The second one is how to overcome the influence of outliers that exist in the original signal vector. For improving the performance of 1D-TP, shapelets transform strategy is introduced.

*3.2. Improved 1D-TP feature exteraction with shapelets transform strategy*

The shapelets transform [25] is a shape-based time series representation approach which could identify the local or global similarity of shape and offer an intuitively comprehensible way of understanding long time series. It could be executed with the following steps as show in Figure 1.

First of all, assuming $Ts$ is the original time series collection, and randomly divide *it* to generate a set of shapelets candidates $W = \{w_{min}, w_{min+1}, \ldots\ldots w_{max}\}$ from the entire sample set including the time series sample $Ts=\{Ts_1, Ts_2, Ts_3 \ldots . Ts_n\}$ and corresponding class labels.

Secondly, calculate the Euclidean distances which are the values between all shapelets candidates $w_i$ and the time series in $Ts$ as Eq.2, and sort them in increasing order to form the set Ds

$$Ds(w_i, Ts_i) = \sum_{i=1}^{t}(w_i - Ts_i)^2 \qquad (2)$$

Next, information gain (IG) [24] defined as Eq.3 and 4 for each candidate in W is calculated, thus their qualities could be assessed

$$H(Ds) = -p(M)\log(p(M)) - p(N)\log(p(N)) \qquad (3)$$

$$IG = H(Ds) - (\frac{|Ds_m|}{|Ds|})H(Ds_m) + \frac{|Ds_n|}{|DS|}H(Ds_n) \qquad (4)$$

$p(M)$ and $p(N)$ respectively represent the proportion of different fault categories $M$ and $N$ in the sample set $Ds$, $Ds_m$ and $Ds_n$ are two subsets in $Ds$.

Finally, shapelets are extracted from the generated candidates by comparing their IG values, and the data set $W_{new}$ could be established, which is composed of the extracted shapelet and arranged in descending order of IG value. The sequence after the shapelets transform is the sub-sequence with higher fault information.

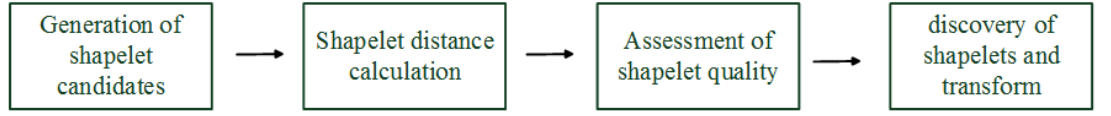

**Figure 1.** diagram of shapelets transform

Through the aforementioned shaplets transform method, the information gain of the local subsequence relative to the original signal is compared, thus the extracted shapelets contain the main information of the original fault signal, and the outliers with little fault information content can be removed.

Thus for the original data, the influence of outliers could be eliminated with shapelets transform, and extract similar local subsequences. Then with the help of ID-TP, the features of the local sub-sequences could be extracted. During the procedure, the standard deviation of the local subsequence after the shapelets transform, which is calculated with Eq.5, is assigned to the parameter of β in ID-TP, here $w_i$ represents the transformed shapelets, $n$ is the number of points in the sequence $W_{new}$:

$$\beta = \sqrt{\frac{\sum_{i=1}^{n}(w_i - \bar{w_i})^2}{n-1}} \tag{5}$$

Thereby the effect of the one-dimensional ternary pattern feature extraction method could be improved. The detailed procedure could be described as Table 2.

**Table 2.** Overview of proposed method

| Feature extraction method with shapelets transform and 1D-TP |
|---|
| 01   Initialization, discovering shapelets |
| 02   Import series dataset |
| 03   *min_ shapelet* (default = 3) |
| 04   *r* (maximum number of shapelets to store, default = 10*size of TS) |
| 05   *quality* (predefined information gain threshold, default = 0.05) |
| 06   calculate the Euclidean distances between shapelets candidates by Eq2 |
| 05   calculate information gain by Eq3 and Eq4 |
| 06   Set shapelets candidates |
| 07   Calculate the standard deviation of parameter *β* as shapelets candidates by Eq 6 |
| 08   switch *Pi* |
| 09       case $P_c < P_i + \beta$ |
| 09           *bainarycode*=1 |
| 10       case $P_c \in (P_i + \beta, P_i - \beta)$ |
| 11           *bainarycode*=0 |
| 12       case $P_c < P_i - \beta$ |
| 13           *bainarycode*=-1 |
| 14       end |
| 15   Separation of positive and negative values; |
| 16   conversion binary values to decimal; |
| 17   end |

Obviously, the improved feature extraction method with 1D-TP and shapelets transform can not only quickly obtain the optimal parameters, but also can filter out abnormal signals which affect fault recognition..With the extracted features obtained from the improved feature extraction strategy, classification methods will be introduced to detect the bearing faults.

## 4.The improved classification strategy

*4.1 XGBoost classifier*

XGBoost is highly efficient classification strategy based on Gradient Boosting Decision Tree(GBDT) which is widely used in biological sciences, and prediction of risks and financial etc[26-29]. Its basic idea is to combine multiple tree models with low classification accuracy to construct a more complex model with relatively high accuracy. The strategy executes iteratively, and in each iteration a new tree used to fit the residual of the previous tree will be generated. Compared with the traditional GBDT, the second derivative and regular term are introduced to make the loss function more accurate and avoid tree overfitting respectively.

In the classification strategy of ensemble learning, the selection of the parameter values of the classifier has an important effect on the performance of the classifier. Table 3 lists the parameters and ranges that can be optimized by the XGBoost classifier. In order to optimize the selection of classifier parameters in XGBoost classifier, this paper introduces artificial bee colony algorithm which has been proven to be more competitive with other population-based algorithms.

**Table 3.** XGBoost classifier related parameters and range

| Paremeter | Range | Meaning |
|---|---|---|
| n_estimators | [1,1000] | Number of trees |
| colsample_bytree | [0,1.0] | subsampling of columns |
| learning_rate | [0.001,0.9] | Step size shrinkage used in update to prevents overfitting |
| min_child_weight | [1,100] | Minimum sum of instance weight (hessian) needed in a child |
| max_depth | [1,15] | Maximum depth of a tree |

*4.2 Improved artificial bee colony algorithm*

ABC algorithm is an optimization algorithm based on the intelligent foraging behaviour of honey bee swarm which consists of three groups: employed bees, onlooker bees, and scout bees. The location of the food source represents the optimal solution to the problem, and its quality could be measured with the value of fitness function.

The traditional algorithm iteratively optimizes from the stage of employed bees, traverses each randomly generated solution vector and updates the solution with neighborhood search as Eq.7. Then, according to the selection probability generated by

Eq.8, the onlooker bee selects one of the generated solutions to execute the subsequent exploration.

$$x_{ik}^* = x_{ik} + rand(Vmin, Vmax)(x_{ik} - x_{jk}) \tag{7}$$

$$P_i = \frac{fit_i}{\sum_{i=1}^{N} fit_i} \tag{8}$$

where $x_{ik}$ represents the solution of the last iteration of the employed bee, $x_{ik}^*$ represents new solution that updated based on $x_{mk}$ as Eq.7, and i, j represent the indices of specific solution in the population, $i,j \in \{1,2,...,N\}$, $i \neq j$. k represents the dimension of the population, $k \in \{1,2,...,D\}$. $rand(-1,1)$ is a random number between [-1,1], $fit_i$ is the fitness value of the objective function, N is the number of food sources which is equal to the number of employed bees. $P_i$ represents the selection probability of each solution vector.

As metioned above, the randomly search or update are executed in the whole solution space and it is easy to fall into the local optimization. In order to overcome the shortage, an adaptive divergence control mechanism is introduced. As shown in Figure 2, the main steps of the improvements are described as follows:

Step 1: Initialize the parameters of ABC, set the number of bee colony, the maximum number of iterations and so on.

Step 2: Divide the whole solution space into $v$ sub-regions who possess the same intervals. And randomly assign weights to different sub-regions as $\varphi\{\theta_1, \theta_2, ..., \theta v\}$.

Step 3: The employee bee searches in each sub-region, generates a new solution $x_{ik}^*$ according to Eq 7 and calculates the fitness value. According to the fitness value calculated after the sub-region search, the weight of the sub-region is scaled up or down, generating a new set of weights.

Step 4: Calculate the selection probability of each sub-region according to Eq 10

$$P_j = \frac{\theta_j}{\sum_{a=1}^{n} \theta_a} \tag{10}$$

Where $j = 1, 2, ..., n$, $P_j$ is the value of probability of selection, $\theta_j$ is the weight value corresponding to the sub-region, and n is the number of all sub-regions.

Step 5: According to the selection probability of each sub-region, observe the bees first search for important sub-regions and select the nectar source according to the greedy strategy.

Step 6: The scout bee judges whether there is a nectar source that needs to be abandoned, and if it exists, it randomly generates a nectar source to replace it.

Step 7: Record the optimal solution until the iterative termination condition is met and output the optimal solution.

The improved artificial bee colony algorithm will search under the guidance of the weights of sub-region every time, and will increase the search effort for the sub-region that may contain the optimal value. This method speeds up the search for the optimal solution, and at the same time helps to avoid the misjudgment of the local optimal.

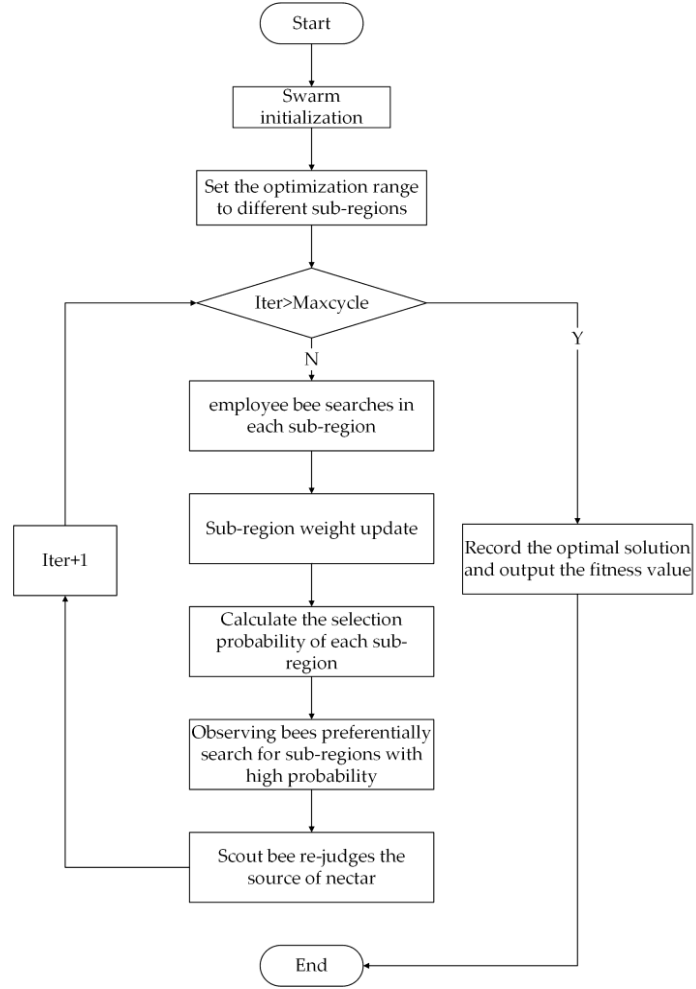

**Figure 2.** Flow chart of adaptive divergence control mechanism

In order to verify the effectiveness of the improved artificial bee colony algorithm optimization ability, five benchmarks from CEC2017 as shown in Table 4 are introduced. The parameters in the bee colony are set as follows: the total number of bee colonies is 200, the maximum number of iterations is 1000, and the search range is the threshold range of the solution of each function.

**Table 4.** The benchmark functions

| Function expression | Search space | Maximum value | Modality |
|---|---|---|---|
| $f_1(x) = \left(20 + \left(x^2 - 10\cos(2\pi x)\right) + \left(y^2 - 10\cos(2\pi y)\right)\right)/\sqrt{2}$ | [-5.12, 5.12] | 118 | Multimodal |
| $f_2(x) = \dfrac{\sin(x)}{x} \Big/ \dfrac{\sin(y)}{y}$ | [-10, 10] | 1 | Unimodal |

| Function expression | Search space | Minimum value | Modality |
|---|---|---|---|
| $f_3(x) = x^2 + y^2$ | [-10, 10] | 0 | Unimodal |
| $f_4(x) = (x\sin(4\pi x) - y\sin(4\pi y + \pi + 1))/2$ | [-1,2] | -1.5 | Unimodal |
| $f_5(x) = \left(20 + \left(x^2 - 10\cos(2\pi x)\right) + \left(y^2 - 10\cos(2\pi y)\right)\right)/\sqrt{2}$ | [-5.12, 5.12] | 0 | Multimodal |

Figure 3 show the fitness values with the traditional ABC and the improved algorithm corresponding to five benchmarks, and Table 5 is the statistics information of the results. It is clear that the improved ABC possesses better convergence performance than ABC for all the benchmark functions, especially for $f_2(x)$ the covergence speed increases by 55.7%. the improved artificial bee colony algorithm can have fewer iterations and higher iteration accuracy. Meanwhile for the optimization accuracy, the resutls corresponding to the improved algorithm possess better performance, where the fitness value obtained by improved algorithm is a 36.2% reduction with respect to ABC for $f_3(x)$.

Thereby the effectiveness of the improved bee colony algorithm could be verified, and further application of the optimization of classifier parameters is possible.

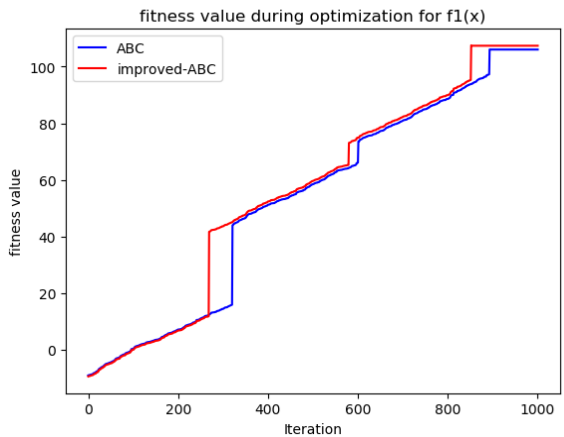

(a)

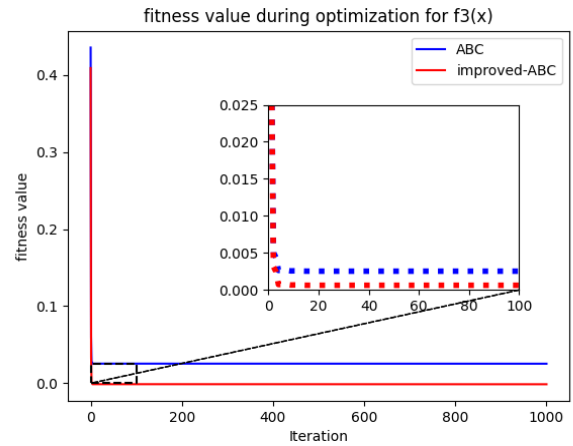

(c)

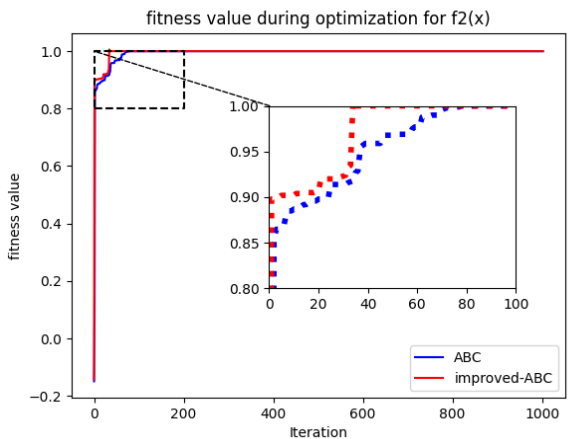

(b)

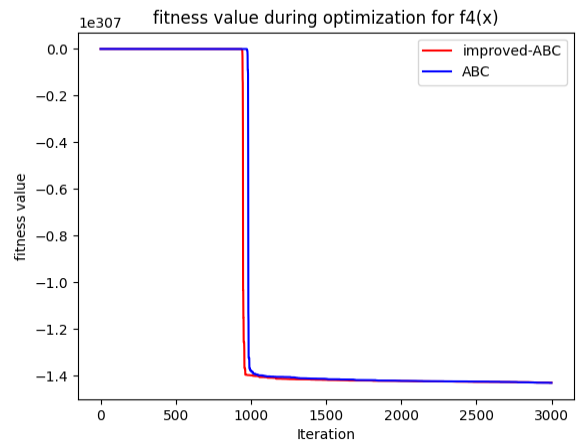

(d)

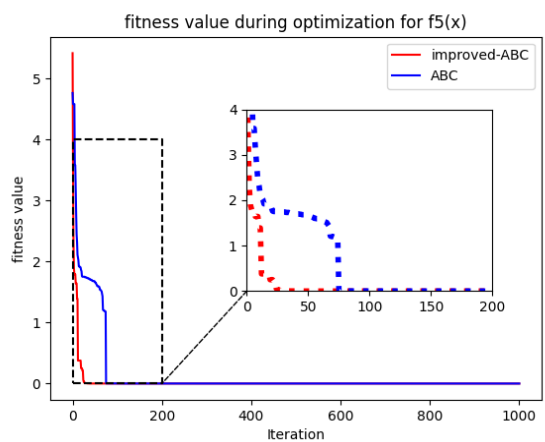

(e)

**Figure 3.** Fitness values of five benchmark functions

**Table 5.** Statistics information of different benchmark functions

| functions | method | Best-F | iterations | Optimal solution |
|---|---|---|---|---|
| $f_1(x)$ | ABC | 117.9375 | 932 | (x=-4.83,y=4.96) |
| | IABC | 117.9402 | 876 | (x=-4.87,y=4.98) |
| $f_2(x)$ | ABC | 1.0000 | 79 | (x=0,y=0) |
| | IABC | 1.0000 | 35 | (x=0,y=0) |
| $f_3(x)$ | ABC | 1.35E-19 | 53 | (x=0.34,y-0.062) |
| | IABC | 0.86E-19 | 37 | (x=0.14,y=0.047) |
| $f_4(x)$ | ABC | -1.4301 | 2983 | (x=1.83,y=-0.93) |
| | IABC | -1.4302 | 2924 | (x=1.95,y=0.95) |
| $f_5(x)$ | ABC | 6.77E-12 | 992 | (x=0.013,y-0.052) |
| | IABC | 0.0000 | 996 | (x=0.0,y=0.0) |

*4.3. XGBoost strategy with improved ABC*

Based on the evaluation of improved ABC in the previous section, a new framework as shown in Figure 4 where the improved bee colony algorithm is applied to optimize the parameters of XGBoost classifier is proposed. The main process is described as Figure 4:

Step 1: Import and preprocess the bearing fault data.
Step 2: Initialize the relevant parameters of ABC and the optimized classifier parameters.
Step 3: Divide the solution spaces into different sub-regions.
Step 4: Execute the search with seperated groups in different sub-regions, and update the weight values of each subregions according to the fitness values of optimal solution searched in that region. Here the the solution is the paramters of XGBoost and the fitness value is the classification accuracy corresponding to each solution.
Step 5: Under the guidance of the weights, onlooker bees and scout bees perform the further exploration.
Step6: After reaching the maximum number of iterations, output the optimized parameters.

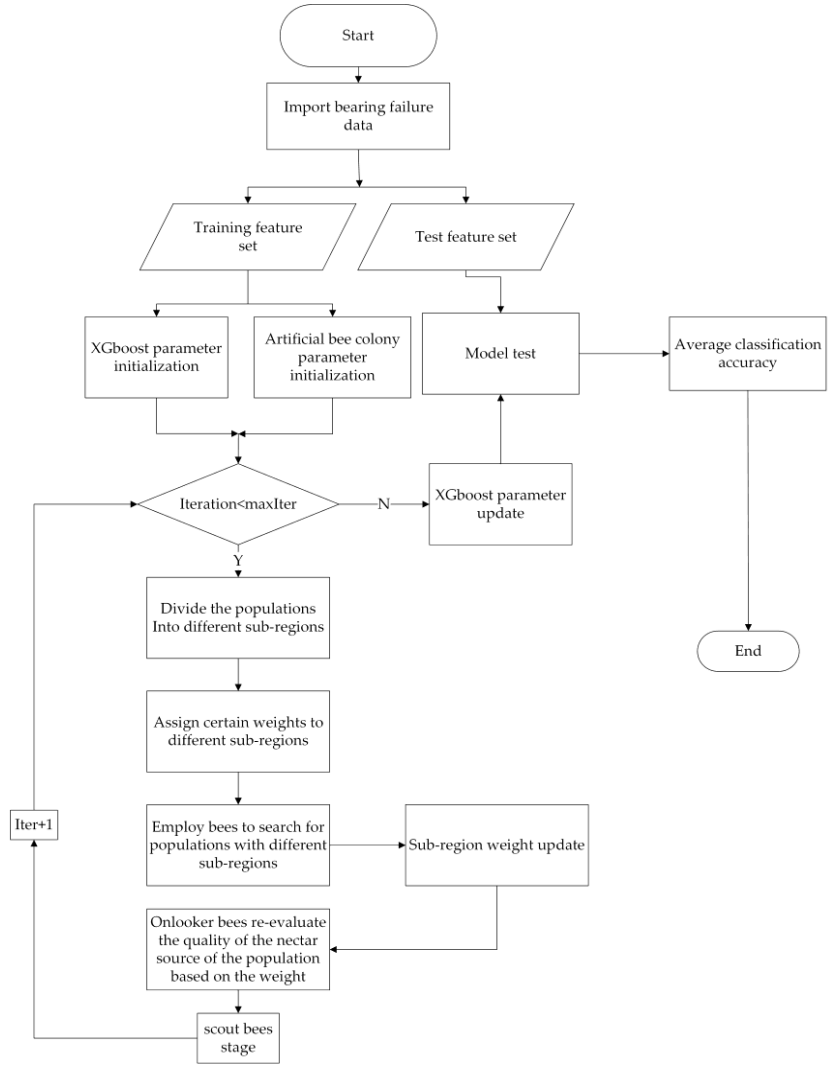

**Figure 4.** Improved bee colony algorithm to optimize XGBoost

## 5. Experiments and Results

With the bearing fault data mentioned in Section 2, the effectiveness of the improved fault detection method is tested in this part.

*5.1 Comparision of feature extraction method*

In order to verify the performance of the improved feature extraction strategy, XGboost classifier whose parameters are selected by experience is introduced. The corresponding values of XGboost are set as: n_estimators=100，learnrate=0.1. Figure 5 shows the confusion matrix with 1D-TP and 1D-TP-ShapeletTransform strategies. The red numbers in the white squares represent the probability of misjudgment of the label. The number of misjudged labels in Figure 5(a) is obviously more than that in Figure 5(b), which means that the improved feature extraction method has better performance. Figure 6 is a histogram of the comparison between the actual number of labels and the number of accurate predictions on the test data set. The number of correct predictions for label 2,

3 and 8 has increased by 10, 10 and 7 respectively, and the number of correct predictions for other labels has also significantly increased.

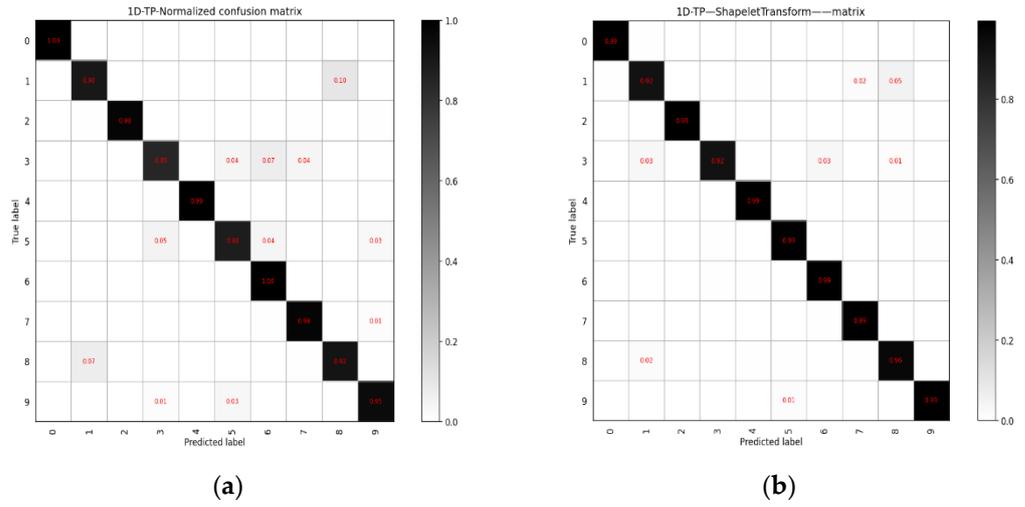

(**a**)                                                       (**b**)

**Figure 5** (a) is the confusion matrix wtih 1D-TP-XGBoost，(b) is the confusion matrix with 1D-TP-ShapeletTransform-XGBoost

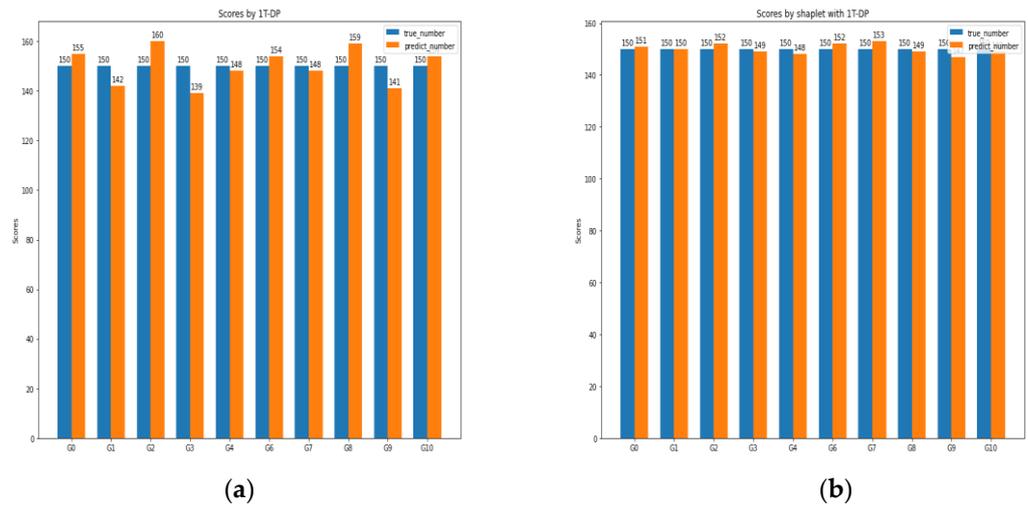

(**a**)                                                       (**b**)

**Figure 6** (a) is the prediction of the test set with 1D-TP-ShapeletTransform method, (b) is the prediction of the test set with 1D-TP

Figure 7 presents the receiver operating characteristic curves (ROC) with two methods. It is clear that the area below ROC curve corresponding to the improved 1D-TP method is larger than the other one, explain that the improved feature extraction method makes the ROC curve of the classifier better.

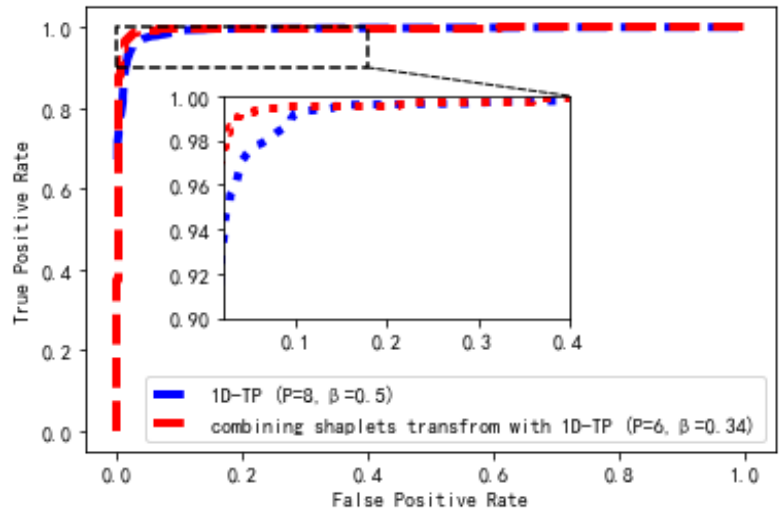

**Figure 7.** Comparison of ROC between the feature extraction method

Other feature extraction method such as FFT are also introduced to compare with with improved 1D-TP strategy, and the results are shown in Figure 8 and Table 6. As the 1D-TP method needs to establish ternary patterns over the entire time series, it takes longer computing time than improved 1D-TP . Compared with the traditional time-domain and frequency-domain feature extraction methods, the error rate was reduced by about 18.75%. The accuracy of the improved feature extraction method is about 97%, which is much higher than other methods.

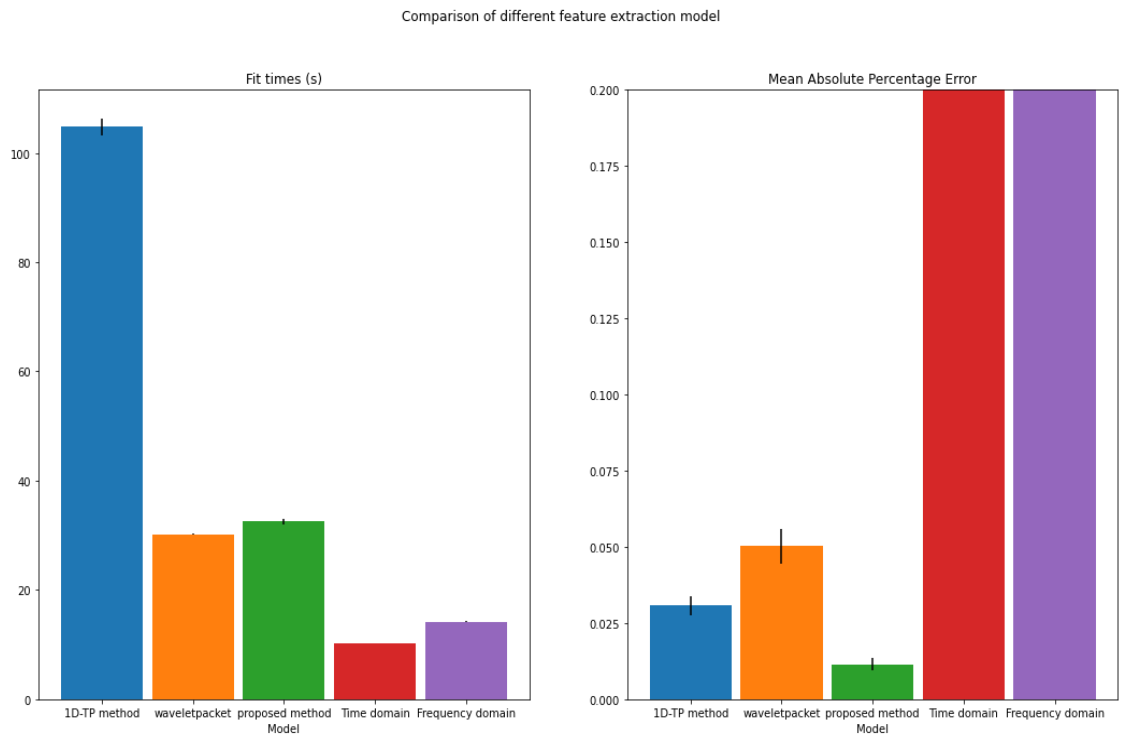

**Figure 8.** Comparison of fit time and error of different feature extraction strategies

**Table 6**. Accuracy under different feature extraction methods.

| Feature extraction method | Accuracy | Time |
|---|---|---|
| shapelets transform with 1D-TP | 97.02% | 36s |
| 1D-TP | 95.03% | 90s |
| Time domain(Root mean square, pulse factor, etc.) | 80.66% | 15s |
| FFT (fast Fourier transform) | 80.08% | 18s |
| Wavelet packet transform | 94.35% | 32s |

*5.2 Comparision of XGBoost parameter optimization*

From the experimental comparison in the previous section, it can be seen that the fault classification model established by Shaplets transform-1D-TP combined with XGBoost has the best performance In order to further improve the classification model, The experiment in this section is based on the improved feature extraction method proposed ,the improved artificial bee colony algorithm proposed in the previous chapter be applied to the optimization of XGBoost parameters in this paper. Table 7 lists the parameters that need to be automatically optimized. During the optimization process, the parameters in the bee colony are set as follows: the total number of bee colonies is 200, the maximum number of iterations is 1000, the range of n_estimators is [1,1000], and the optimization range of learning rate is [0.01,1].

**Table 7.** The parameters of XGBoost

| parameters | range | Meaning | Optimal parameter setting |
|---|---|---|---|
| **n_estimators** | [1,1000] | Number of trees | 876 |
| **learning rate** | [0.01 1] | Step size shrinkage used in update to prevents overfitting | 0.26 |

In this experiment, the classification accuracy is used as the objective function, and the iterative optimization curve is compared as shown in Figure 9. Figure 9 and Table 8 show that the iterative optimization curve of the improved artificial bee colony algorithm finds the optimal value of accuracy faster and converges, and improves the accuracy to about 98.60%.

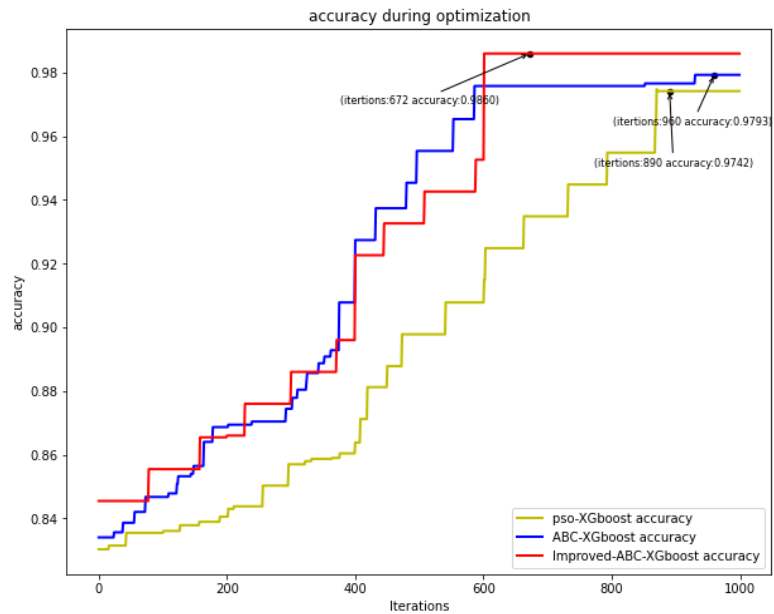

**Figure 9.** Iterative optimization curve comparison

**Table 8.** Model results

| Model | Accuracy | Iteration |
|---|---|---|
| **PSO-XGBoost** | 0.9742 | 890 |
| **ABC-XGBoost** | 0.9793 | 960 |
| **IABC-XGBoost** | 0.9860 | 672 |

## 6. Conclusions

In this paper, for solving the problem of bearing fault detection, a novel feature extraction method is introduced based on the traditional 1D-TP. In order to obtain the optimal parameters which is usually selected by trial and error and overcome the influence of outliers in original signals, the shapelets transform method is proposed to optimize 1D-TP. Meanwhile, an improved ABC algorithm is created and introduced to optimize the performance of XGBboost classifier. The improved algorithms of feature extraction and classification are applied to the problem of bearing fault detection, and results show the effectiveness of the proposed strategies.

In future work, more advanced machine learning algorithms need be applied to improve detection accuracy, and multi-core parallel computing will be used to accelerate the speed of optimization procedure.

**Supplementary Materials:** The following are available online at www.mdpi.com/xxx/s1, Figure 1: diagram of shapelets transform ,Figure 2: Flow chart of adaptive divergence control mechanism, Figure 3: Fitness values of five benchmark functions, Figure 4: Improved bee colony algorithm to optimize XGBoost, Figure 5: Confusion matrix comparison Figure 6: Comparison of sample classification results ,Figure 7: Comparison of ROC between the feature extraction method ,Figure 8: Comparison of fit time and error of different feature extraction strategies,Figure 9: Iterative optimization curve comparison,Table 1: Establishment of sample data , Table 2: Overview  of proposed method,Table 3: XGBoost classifier related parameters and range ,Table 4: The benchmark functions,Table 5: Statistics information of different benchmark functions ,Table 6: Accuracy under different feature extraction methods,Table 7: The parameters of XGBoost ,Table 8: Model results.

**Author Contributions:** Conceptualization, Haiquan Wang and Wenxuan Yue; Methodology, Haiquan Wang and Wenxuan Yue; software, Haiquan Wang and Wenxuan Yue; Validation, Shengjun Wen and Xiaobing Xun; Formal analysis , Menghao Su and Shanshan Zhang; Investigation, Panpan Du; Data curation, HaiquanWang and WenxuanYue; writing—original draft preparation, Haiquan Wang and WenxuanYue; All authors have read and agreed to the published version of the manuscript.

**Funding:** This work was supported by the Training Program for Young Teachers in Universities of Henan Province, grant number 2020GGJS137; the Technical Guidance Project of Textile Industry Federation, grant number 2020114; the NSFC-Zhejiang Joint Fund for the Integration of Industrialization and Informatization, grant number U1709215; the Zhejiang Province Key R&D projects, grant number No.2019C03104; the Fundamental Research Funds of Zhongyuan University of Technology, grant number K2019YY005; Henan Province Science and Technology R&D projects, grant number 202102210135, 212102310547 and 212102210080; National Nature Science Foundation of China, grant number U1813201.

**Data Availability Statement:** The data set established in this experiment is provided by Bearing Data Center in Case Western Reserve University.

**Conflicts of Interest:** The authors declare no conflict of interest.